\documentclass[preprint,showpacs,preprintnumbers,amsmath,amssymb]{revtex4}

\usepackage{graphicx}
\usepackage{dcolumn}
\usepackage{bm}

\begin{document}

\preprint{}

\title{Joint Hilbert-Schmidt Determinantal  Moments of Product Form for Two-Rebit and Two-Qubit and Higher-Dimensional Quantum Systems}

\author{Paul B. Slater}%
\email{slater@kitp.ucsb.edu}
\affiliation{%
University of California, Santa Barbara, CA 93106\\
}%
\date{\today}

\begin{abstract}
We report formulas 
for the joint moments of the determinantal products 
$(\det{\rho})^{k} (\det{\rho^{PT}})^{\kappa}$ ($k=0, 1, 2,\ldots,N$; $\kappa = 1, \ldots, 12$) of Hilbert-Schmidt (HS) probability distributions over the  generic two-rebit and  two-qubit density matrices  
$\rho$ ($\kappa =1,  \ldots, 4$). Here $PT$ denotes the partial transposition operation of quantum-information-theoretic central importance. Each formula is the product of the expression for the 
HS moments of $(\det{\rho})^{k}$, $k=0,1,2,\ldots,N$--special cases of results of Cappellini, Sommers and {\.Z}yczkowski ({\it Phys. Rev. A} {\bf{74}}, 062322 (1996))--and an adjustment factor. 
The factor is a {\it biproper} rational function, with its numerators and denominators both being $3 \kappa$-degree polynomials in $k$. We infer the structure that the denominators follow for arbitrary $\kappa$ in both the two-rebit and two-qubit cases, and the six leading-order coefficients of 
$k$ of the numerators in the two-rebit scenario. We also commence an analogous investigation of generic rebit-retrit and qubit-qutrit systems. This research was motivated, in part, by the objective of using the computed moments to well reconstruct the HS probabilities over the determinant of $\rho$ and of its partial transpose, and to ascertain--at least to high accuracy--the associated (separability) probabilities of "philosophical, practical and physical" interest that  
$(\det{\rho^{PT}})>0$.
\end{abstract}


\maketitle
We begin our investigation into certain statistical aspects of the "geometry of quantum states" \cite{ingemarkarol,BookReview} by noting the two following special cases--which we will extend below--of the general formulas \cite{csz}[eq. (3.2)]:
\begin{equation} \label{firstold}
\langle |\rho|^k \rangle_{2-rebit/HS}=945 \Big( 4^{3-2 k} 
\frac{ \Gamma (2 k+2) \Gamma (2 k+4)}{\Gamma (4 k+10)} \Big)
\end{equation}
and 
\begin{equation} \label{secondold}
\langle |\rho|^k \rangle_{2-qubit/HS}=108972864000 \frac{ \Gamma (k+1) \Gamma (k+2) \Gamma (k+3) \Gamma
   (k+4)}{\Gamma (4 (k+4))},
\end{equation}
$k=0, 1, 2,\dots$
The bracket notation $\langle \rangle$ is employed by us to denote expected value, 
while $\rho$ indicates a generic
(symmetric) two-rebit or generic (Hermitian) two-qubit ($4 \times 4$) density matrix. The expectation is taken with respect to the probability distribution determined by the 
Hilbert-Schmidt/Euclidean/flat metric on either the 9-dimensional space of generic two-rebit or 15-dimensional space of generic two-qubit systems \cite{szHS,ingemarkarol}.

We report below {\it sixteen} (twelve two-rebit and four two-qubit) non-trivial extensions of these formulas, involving now in addition to $|\rho|$, the quantum-theoretically important determinant $|\rho^{PT}|$ of the partial transpose of $\rho$. (The nonnegativity 
of $|\rho^{PT}|$--by the celebrated Peres-Horodeccy results \cite{asher,michal,augusiak}--constitutes a necessary and sufficient condition for separability/disentanglement, when $\rho$ is either a $4 \times 4$ or $6 \times 6$ density matrix.) At this point of our presentation, we note that three of these extensions are expressible--incorporating on their right-hand sides the two formulas above--as 
\begin{equation} \label{firstnew}
\langle |\rho|^k |\rho^{PT}| \rangle_{2-rebit/HS}= 
\frac{(k-1) (k (2 k+11)+16)}{32 (k+3) (4 k+11) (4 k+13)} \langle |\rho|^k \rangle_{2-rebit/HS},
\end{equation}
\begin{equation} \label{secondnew}
\langle |\rho|^k |\rho^{PT}|^2 \rangle_{2-rebit/HS}= 
\frac{k (k (k (k (4 k (k+12)+203)+368)+709)+2940)+4860}{1024 (k+3) (k+4)
   (4 k+11) (4 k+13) (4 k+15) (4 k+17)} \langle |\rho|^k \rangle_{2-rebit/HS}
\end{equation}
and
\begin{equation} \label{thirdnew}
\langle |\rho|^k |\rho^{PT}| \rangle_{2-qubit/HS}= 
\frac{k (k (k+6)-1)-42}{8 (2 k+9) (4 k+17) (4 k+19)} \langle |\rho|^k \rangle_{2-qubit/HS}.
\end{equation}

These three formulas were, first, established by "brute force" computation--that is  calculating the first ($k=0, 1, 2,\ldots,15$ or so) instances, then employing
the Mathematica command FindSequenceFunction, and verifying any formulas
generated on still higher values of $k$. ({\it Initially}, although we had the specific values of 
$\langle |\rho|^k |\rho^{PT}|^3 \rangle_{2-rebit/HS}$ for $k=0,\ldots,13$, and similarly for $\langle |\rho|^k |\rho^{PT}|^4 \rangle_{2-rebit/HS}$, we were not able to determine, in the same manner, 
encompassing expressions for them.)

As a special case ($k=1$) of (\ref{firstnew}), we obtain the rather remarkable moment result, zero, already reported in \cite{HSorthogonal}. The immediate interpretation of this finding is that for the generic two-rebit systems, the two determinants
$|\rho|$ and $|\rho^{PT}|$ comprise a pair of nine-dimensional {\it orthogonal} polynomials \cite{dunkl2,dumitriu,griffithsspano} with respect to Hilbert-Schmidt measure. (C. Dunkl has kindly pointed out that orthogonality here does not imply zero {\it correlation}.) 
In addition to this first ($k=1$) 
HS zero-moment of the ("equally-mixed") product variable
$|\rho| |\rho^{PT}|$ in the two-rebit case, we had been able to compute its higher-order  moments, $k =2,\ldots,6$. (The results for $k=2$ can be obtained by direct application of (\ref{secondnew}). The feasible range of the variable is 
$|\rho| |\rho^{PT}| \in [-\frac{1}{110592},\frac{1}{256^2}]$--the lower bound of which$-\frac{1}{110592} =- 2^{-12} 3^{-3}$ we determined by analyzing a general convex combination of a Bell state and the fully-mixed state.) 

These five further moments of  $|\rho| |\rho^{PT}|$,  $k =2,\ldots,6$, are all rational numbers. T
If we take the ratios of these first six moments
of $|\rho| |\rho^{PT}|$ to the first six {\it even} moments given by (\ref{firstold}), that is the values $\langle |\rho|^{2 k} \rangle_{2-rebit/HS}$, $ k=1,\ldots,6$, we obtain the 
rather succinct sequence,
\begin{equation} \label{sequence}
\frac{\langle (|\rho| |\rho^{PT}|)^k \rangle_{2-rebit/HS}}{\langle |\rho|^{2 k} \rangle_{2-rebit/HS}} =\{ 0, \frac{77}{54}, \frac{24}{55}, \frac{209}{175}, \frac{598}{833},  \frac{3929}{3724} \}
\end{equation}
\begin{displaymath}
\approx \{0, 1.425926, 0.4363636, 1.194286, 0.7178872, 1.055048\}.
\end{displaymath}
(As to the two-{qubit} counterpart of this sequence, 
we had so far only been able to compute its very first term--turning out, quite remarkably, 
to be the negative value $-\frac{3}{2}$.) 

Since these ratios (\ref{sequence}) are so comparatively simple, it suggested to us that we might be more able to progress in a series of analyses \cite{ZHSL,advances,ratios,maxconcur4,JMP2008,slater833,slaterJGP2,slaterPRA2} (mainly devoted to the determination of {\it separability} probabilities), by making our initial goal
the computation of  these 
ratios for still higher-order moments--rather than the direct computation of the very small values, having lengthy multi-digit denominators, of the moments $\langle (|\rho| |\rho^{PT}|)^k \rangle_{2-rebit/HS}$  themselves. 
(In \cite{HSorthogonal}[eqs, (33)-(41)], we were able to report and analyze the first nine moments $\langle |\rho^{PT}|^k \rangle$--the first two of which, $-\frac{1}{858}$ and $\frac{27}{2489344}$, can be obtained by directly setting $k=0$ in (\ref{firstnew}) and (\ref{secondnew}), respectively. However, to this point, we have not found any associated similarly compact sequences of moment ratios, as above.)

\begin{figure}
\includegraphics{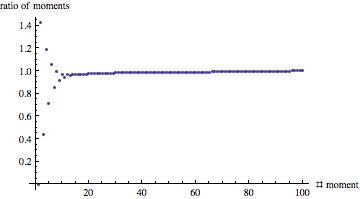}
\caption{\label{fig:RebitListPlot}The six-member 
two-rebit HS exact moment ratio 
sequence (\ref{sequence}), supplemented by its numerical continuation, using extended-precision (60-digit) arithmetic. Three hundred and twenty million random density matrices were employed.}
\end{figure}

\begin{figure}
\includegraphics{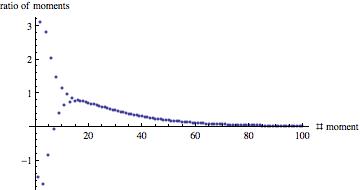}
\caption{\label{fig:QubitListPlot}The two-qubit analogue of the two-rebit sequence depicted in 
 Fig.~\ref{fig:RebitListPlot}, with only the first member ($-\frac{3}{2}$) having initially been exactly known, and the next ninety-nine 
computed numerically, using extended-precision (60-digit) arithmetic. Twenty-four million random density matrices were employed.}
\end{figure}

Accordingly, in Fig.~\ref{fig:RebitListPlot}, we display the sequence of ratios 
$\frac{\langle (|\rho| |\rho^{PT}|)^k \rangle_{2-rebit/HS}}{\langle |\rho|^{2 k} \rangle_{2-rebit/HS}}$, $k =1, \ldots, 100$, the first six members of which have been exactly calculated, as noted above, and the rest through extended-precision (60-digit) numerical computations. (Simply as an indicator of accuracy of these computations, 
the {\it numerical} estimates yielded by this procedure of the six-member {\it exact} sequence (\ref{sequence}) were $\{-0.0002052822, 1.426286, 0.4359643, 1.194784, 0.7175908, 1.055326\}$.) Similarly, in Fig.~\ref{fig:QubitListPlot}, we display the (quite differently-behaving) 
two-qubit sequence of moment ratios 
$\frac{\langle (|\rho| |\rho^{PT}|)^k \rangle_{2-qubit/HS}}{\langle |\rho|^{2 k} \rangle_{2-qubit/HS}}$, $k =1, \ldots, 100$, only the first member of which, -$\frac{3}{2}$, we were initially able to exactly compute, and the remaining ninety-nine, numerically, using extended-precision. 

If we are, at some point, in the course of these extended analyses, able to develop
formulas explaining the {\it full} sequences of ratios in the two-rebit and two-qubit cases, we should be able to reconstruct the Hilbert-Schmidt {\it univariate} probability distributions over the product variable $|\rho| |\rho^{PT}|$ (cf. \cite{csz}[Figs. 2-4]). From such reconstructed distributions, HS {\it separability} probabilities should be determinable to high accuracy.

\begin{figure}
\includegraphics{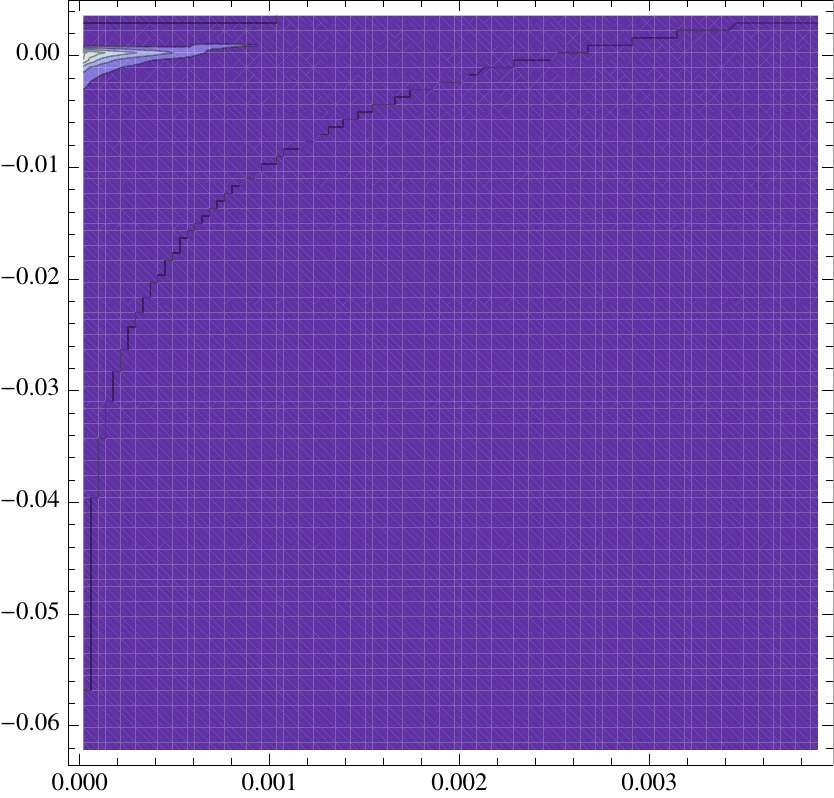}
\caption{\label{fig:JointRebit}Contour plot of the joint Hilbert-Schmidt probability distribution of $|\rho|$ (horizontal axis) and $|\rho^{PT}|$ in the two-rebit case. Larger values appear lighter. The variable ranges are $|\rho| \in [0,\frac{1}{256}]$ and $|\rho^{PT}| \in [-\frac{1}{16},\frac{1}{256}]$. One billion random density matrices were employed.}
\end{figure}

\begin{figure}
\includegraphics{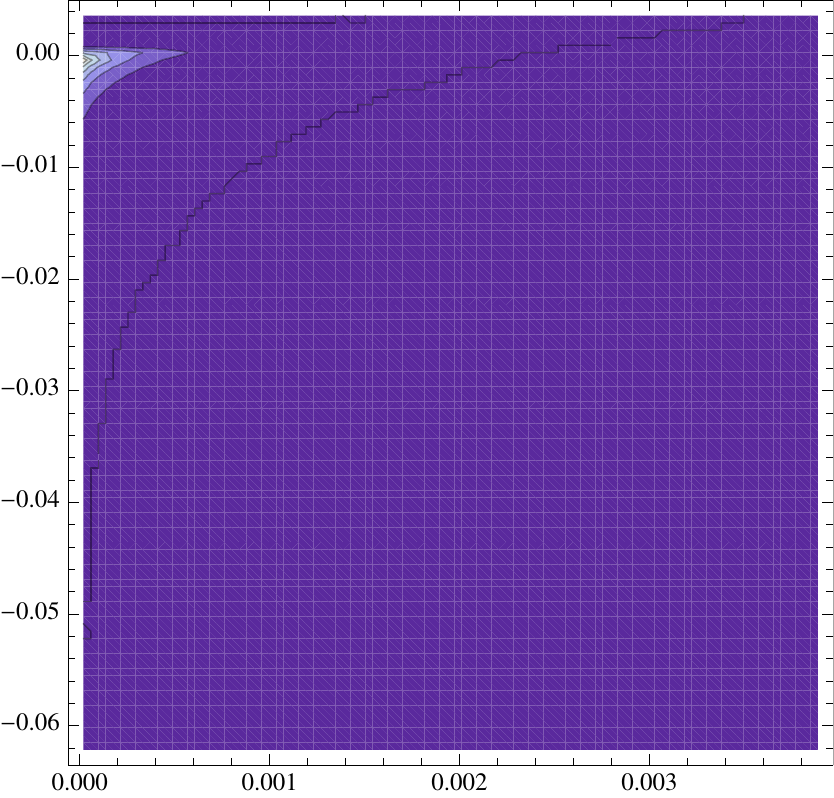}
\caption{\label{fig:JointQubit}Contour plot of the joint Hilbert-Schmidt probability distribution of $|\rho|$ (horizontal axis) and $|\rho^{PT}|$ in the two-qubit case. Six hundred million random density matrices were employed.}
\end{figure}

\begin{figure}
\includegraphics{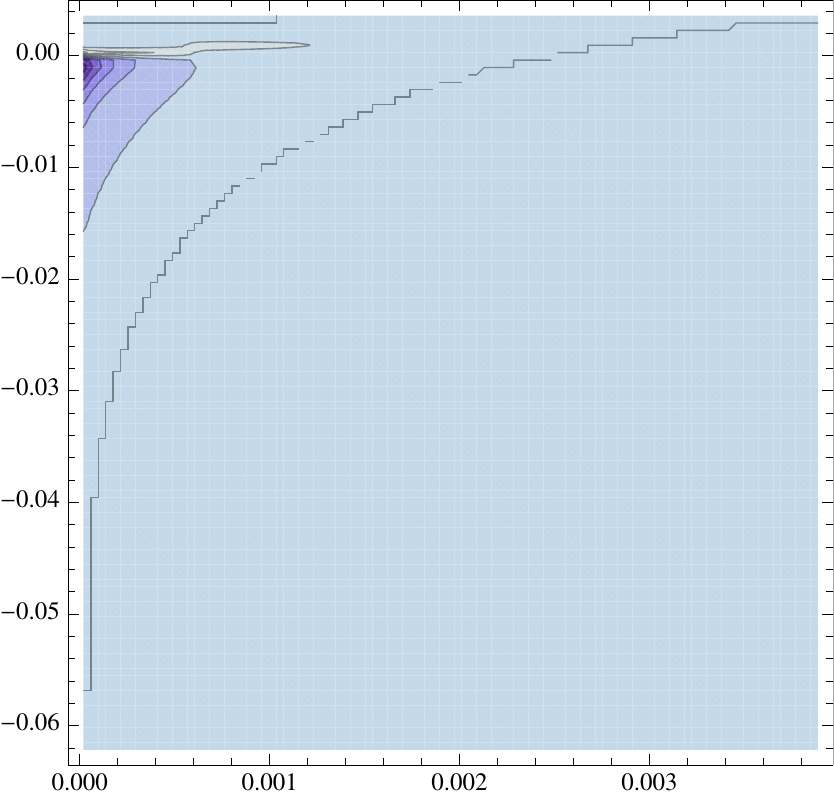}
\caption{\label{fig:JointDiff}Difference obtained by subtracting 
the two-qubit HS probability distribution in Fig.~\ref{fig:JointQubit} from the two-rebit probability distribution in Fig.~\ref{fig:JointRebit}. Darker colors indicate more negative values.}
\end{figure}

For the further edification of the reader, we present in Fig.~\ref{fig:JointRebit}  a contour plot of the joint Hilbert-Schmidt 
(bivariate) probability distribution of $|\rho|$ and $|\rho^{PT}|$ in the two-rebit 
case, and in  Fig.~\ref{fig:JointQubit}, its two-qubit analogue. (A colorized grayscale output is employed, in which larger values appear lighter.) In 
Fig.~\ref{fig:JointDiff} is displayed the {\it difference} obtained by subtracting the second (two-qubit) distribution from the first (two-rebit) 
distribution.
(The black curves in all three contour plots appear to be attempts by Mathematica to establish the nonzero-zero probability boundaries--which, it would, of course, be of interest to explicitly determine, if possible--of the joint domain of $|\rho|$ and $|\rho^{PT}|$.)

These  last three figures are based on  Hibert-Schmidt sampling (utilizing Ginibre ensembles \cite{csz}) of random density matrices, using $10,000=100^2$ bins. In regard to the two-qubit plot, 
K. {\.Z}yzckowski informally wrote:
"A high peak in the upper corner means that: a) a majority of the entangled states is 'little entangled' (small $det(\rho^T)$) or rather, they are 'close' to the boundary of the set, so
one eigenvalue is close to zero, and the determinant is small; 
b) as $det(\rho)$ is also small, it means that these entangled states
live  close to the boundary of the set of all states (at least one eigenvalue is very small), 
but this is very much consistent with the observation that the center of the convex body of the 2-qubit states is separable  (so entangled states have to live  'close' to the boundary).
Similar reasoning has to hold in the real case as well."

At a later point in our investigation, we realized that we might make further
progress--despite limitations on the number of moments we could explicitly compute--by exploiting the evident pattern followed by our newly-found formulas 
(\ref{firstnew}) and (\ref{secondnew})--in particular, the structure in 
their
denominators. This encouragingly proved to be the case, as we were able
to establish that 
\begin{equation} \label{fourthnew}
\langle |\rho|^k |\rho^{PT}|^3 \rangle_{2-rebit/HS}= 
\frac{A_3}{B_3} \langle |\rho|^k \rangle_{2-rebit/HS},
\end{equation}
where
\begin{equation}
A_3=8 k^9+180 k^8+1674 k^7+8559 k^6+29493 k^5+84291 k^4+136801 k^3-401334
   k^2-2516616 k-3612816
\end{equation}
and 
\begin{equation} 
B_3=32768 (k+3) (k+4) (k+5) (4 k+11) (4 k+13) (4 k+15) (4 k+17) (4 k+19) (4
   k+21).
\end{equation}

So, it is now rather evident that we can write for general non-negative integer $\kappa$,
\begin{equation} \label{generalRebit}
\langle |\rho|^k |\rho^{PT}|^\kappa \rangle_{2-rebit/HS}= 
\frac{A_\kappa}{B_\kappa} \langle |\rho|^k \rangle_{2-rebit/HS},
\end{equation}
where both the numerator $A_\kappa$ and the denominator $B_\kappa$ are $3 \kappa$-degree polynomials (thus, forming a "biproper rational function" \cite{chou}) in $k$ (the leading coefficient of $A_\kappa$ being $2^\kappa$), and
\begin{equation} \label{denominator}
B_\kappa=128^{\kappa } (k+3)_{\kappa } \left(2 k+\frac{11}{2}\right)_{2 \kappa },
\end{equation}
where the Pochhammer symbol $(x)_n \equiv \frac{\Gamma(x+n)}{\Gamma(x)} = x (x+1)\ldots(x+n-1)$ is employed.
Further still, moving upward to the next level ($\kappa=4$), we have determined that
\begin{equation} \label{fifthnew}
\langle |\rho|^k |\rho^{PT}|^4 \rangle_{2-rebit/HS}= 
\frac{A_4}{B_4} \langle |\rho|^k \rangle_{2-rebit/HS},
\end{equation}
where
\begin{equation}
\hat{A_4}= 16 k^{12}+576 k^{11}+9112 k^{10}+84496 k^9+525681 k^8+2389416 k^7+7805462
   k^6+13904508 k^5+
\end{equation}
\begin{displaymath}
+6212189 k^4+166748972 k^3+1636873812 k^2+5496485760
   k+6610161600,
\end{displaymath}
and $B_4$ is given by (\ref{denominator})  with $\kappa=4$. 
The real part of one of the roots of  $A_4$ is 2.999905, suggesting to us some possible interesting asymptotic behavior of the roots of these numerators, $\kappa \rightarrow \infty$. In our previous related study \cite{HSorthogonal}[sec. II.B.2], we were also able to discern the general structure that the {\it denominators} of  certain "intermediate [rational] functions" used in computing the (univariate) moments of $\langle \rho^{PT}|^\kappa \rangle_{2-rebit/HS}$, $\kappa=1,\ldots,9$ followed.

From our four new two-rebit moment results (\ref{firstnew}), (\ref{secondnew}), (\ref{fourthnew}) and (\ref{fifthnew}),  we see that the {\it constant} terms in the $3 \kappa$-degree numerator $A_\kappa$
are $-16, 4860, -3612816$ and $6610161600$ for $\kappa=1, 2, 3, 4$. Since we had previously computed \cite{HSorthogonal}[eqs, (33)-(41)]
the moments of $\langle |\rho^{PT}|^\kappa \rangle_{2-rebit/HS}$, 
$\kappa=1,\ldots,9$, we are also able to determine the next five members
of this sequence $\{-16, 4860, -3612816, 6610161600 \}$. However, no general rule for this sequence, which would directly allow us to obtain a formula for $\langle |\rho^{PT}|^\kappa \rangle_{2-rebit/HS}$, has, to our disappointment, yet emerged for them.
(With such a rule, we could address the separability probability question
through the reconstruction of a {\it univariate} probability distribution.)

A simple algebraic exercise involving (\ref{firstold}) shows that if we multiply the conversion factor
\begin{equation}
c= \frac{\sqrt{\pi } \Gamma (2 \kappa +4) \Gamma (8 \kappa +10)}{8 \Gamma
   \left(2 \kappa +\frac{5}{2}\right) \Gamma (4 \kappa +2) \Gamma (4
   \kappa +10)}
\end{equation}
by the rational function
factors $\frac{A_\kappa}{B_\kappa}$ found above that applied to 
$\langle |\rho|^\kappa  \rangle_{2-rebit/HS}$ yield 
$\langle (|\rho| |\rho^{PT}|)^\kappa \rangle_{2-rebit/HS}$, we obtain the 
$\kappa$-member of the sequence of moment ratios (\ref{sequence}).
Since the members of this sequence appear (Fig.~\ref{fig:RebitListPlot}) 
to asymptotically approach 1 (our numerical estimate for the 100-{\it th} term is 1.001542), it would seem that the conversion factor
$c$ and $\frac{B_\kappa}{A_\kappa}$ asymptotically approach one another.

Certainly, it would be of interest to conduct analyses parallel to those reported above for metrics of quantum-information-theoretic interest other than the Hilbert-Schmidt, such as the Bures (minimal monotone) metric 
\cite{szBures,ingemarkarol}. The computational challenges involved, however, might, at least in certain respects, be even more substantial.

At this stage of our research, after posting the results above as a preprint, Charles Dunkl detailed a computational proposal that 
he had outlined to us somewhat earlier. The attractive feature of this proposal would be that it would--holding the exponent $\kappa$ of 
$|\rho^{PT}|$ fixed--be able to compute the adjustment factors for general $k$, rather than having to do so for sufficient numbers of individual members of the sequence $k=1,\ldots, N$, to be able to successfully apply the Mathematica command FindSequenceFunction, as had been our strategy beforehand. The proposal of Dunkl (see Appendix) involved parameterizing $4 \times 4$ density matrices in terms of their Cholesky decompositions. The parameters (ten in number for the two-rebit case and sixteen for the two-qubit case) would be viewed as points on the surface  of a  unit (due to the trace requirement)  
10-sphere or 16-sphere. The squares of the points lie in a simplex.
One can then employ the corresponding Dirichlet distributions over the simplices to determine the corresponding expected values (joint moments).
(A further facilitating aspect here is that both $|\rho|$ and the jacobian for the transformation to Cholesky variables are simply monomials.)
Using this approach, we were able to extend our single ($\kappa=1$) two-{\it qubit} result (\ref{thirdnew}) to the $\kappa=2$ case,
\begin{equation} \label{secondnewqubit}
\langle |\rho|^k |\rho^{PT}|^2  \rangle_{2-qubit/HS}= 
\end{equation}
\begin{displaymath}
\frac{k (k (k (k (k (k+15)+67)+45)+220)+4260)+10944}{64 (2 k+9) (2 k+11)
   (4 k+17) (4 k+19) (4 k+21) (4 k+23)} \langle |\rho|^k \rangle_{2-qubit/HS}.
\end{displaymath}

Additionally, in the following several arrays, we show ($\kappa=1,\ldots,12$) column-by-column, the 
$(3 \kappa +1)$ coefficients of the numerator polynomials in ascending order--the entries in the first row corresponding to the constant terms,\ldots--in the two-rebit case.
For the cases $\kappa=1,\ldots,6$, 
\begin{equation} \label{CoefficientArray1}
\left(
\begin{array}{cccccc}
 -16 & 4860 & -3612816 & 6610161600 & -23680812672000 &
   147885533254368000 \\
 5 & 2940 & -2516616 & 5496485760 & -21644930613600 & 144374531813568000
   \\
 9 & 709 & -401334 & 1636873812 & -7755993054000 & 58524043784903280 \\
 2 & 368 & 136801 & 166748972 & -1199508017652 & 11977854861441312 \\
 - & 203 & 84291 & 6212189 & -4378482660 & 1052189083196640 \\
 - & 48 & 29493 & 13904508 & 29246867605 & -30302414250528 \\
 - & 4 & 8559 & 7805462 & 7876634465 & -6899036908859 \\
 - & - & 1674 & 2389416 & 2649513956 & 3583820785224 \\
 - & - & 180 & 525681 & 883461210 & 1632448582425 \\
 - & - & 8 & 84496 & 219916945 & 477741210624 \\
 - & - & - & 9112 & 40679505 & 118164517947 \\
 - & - & - & 576 & 5660714 & 23817008856 \\
 - & - & - & 16 & 575800 & 3786901675 \\
 - & - & - & - & 40000 & 469728096 \\
 - & - & - & - & 1680 & 44685468 \\
 - & - & - & - & 32 & 3143808 \\
 - & - & - & - & - & 153360 \\
 - & - & - & - & - & 4608 \\
 - & - & - & - & - & 64
\end{array}
\right)
\end{equation}
for $\kappa=7, 8$
\begin{equation}
\left(
\begin{array}{cc}
 -1478171868716632320000 & 22213289955937152264192000 \\
 -1513452218833263744000 & 23581102260458975345971200 \\
 -666431742762610272000 & 11028809359452120997904640 \\
 -159042866967850127040 & 2917571120318305111773312 \\
 -20322562269329104680 & 456603746675135611725072 \\
 -810132874858021428 & 36986782703412677565888 \\
 132995097356746814 & 46803055033262612808 \\
 21392495012420189 & -261889976997036533208 \\
 2726092518763299 & -9109477993443740087 \\
 1004313607984511 & 4010582339036286008 \\
 339532730040063 & 1188217816129015268 \\
 85862223778653 & 313924542711356232 \\
 17623087217813 & 78095538742712398 \\
 2989984893665 & 16444648671941448 \\
 414832485207 & 2889993079367548 \\
 46447024562 & 425775247755632 \\
 4134192972 & 52413983854433 \\
 285783624 & 5342837634528 \\
 14766416 & 445432573168 \\
 535136 & 29856957312 \\
 12096 & 1568760928 \\
 128 & 62140928 \\
 - & 1741568 \\
 - & 30720 \\
 - & 256
\end{array}
\right)
\end{equation}
for $\kappa=9$.
\begin{equation}
\left(
\begin{array}{cc}
 -478312140758259028431974400000  \\
 -522398952485281283922478080000  \\
 -255802623523478150875233408000  \\
 -72876271183514797204357862400  \\
 -12968985411931489708456120320  \\
 -1391118724430074132545941424  \\
 -65460659627903052772432944  \\
 3455511572907702205749432  \\
 652248751783303606044024 \\
 25126439305837999036933  \\
 1467731023076566955385  \\
 1183997600031628820142 \\
 383843332752788856636  \\
 92094572188997438886  \\
 19032858442918448262  \\
 3399681068905043760 \\
 519692040800245908 \\
 67749101821899549  \\
 7505505996250329  \\
 702038674797138 \\
 54914983170480 \\
 3545548122432 \\
 185537763936  \\
 7665748416  \\
 240406272 \\
 5370624  \\
 76032 \\
 512 
\end{array}
\right)
\end{equation}
for $\kappa=10$
\begin{equation}
\left(
\begin{array}{c}
 14207139636708483077514789888000000 \\
 15876402339932769370618323302400000 \\
 8060699595710982594644066482560000 \\
 2429195772959924259673598495040000 \\
 473246195907047274024301397419200 \\
 59861458630180990183509791530560 \\
 4410689910607254768908441542704 \\
 77978523981901390446515581440 \\
 -17699546304068611844956276620 \\
 -1448009072942187597584330860 \\
 23679691892051020702374101 \\
 12657804955771967753837040 \\
 2173418236785245059311455 \\
 535603226600605770029160 \\
 130916289625347750649830 \\
 27174211939515057439440 \\
 4857622320912622046130 \\
 755231957874779338260 \\
 101981982088141404465 \\
 11919937134729947040 \\
 1201281021189026955 \\
 103825857120570960 \\
 7638091703520180 \\
 473427510762240 \\
 24386539757280 \\
 1024463577600 \\
 34174797696 \\
 870092800 \\
 15865600 \\
 184320 \\
 1024
\end{array}
\right)
\end{equation}
for  $\kappa=11$,
\begin{equation}
\left(
\begin{array}{c}
 -564438487027912271386331788873728000000 \\
 -642782791049172077649575252852121600000 \\
 -336051652895031776265874186953960960000 \\
 -105851053751774412125536281940889088000 \\
 -22067056295744606471446676944123680000 \\
 -3120323479792255728315846426979831680 \\
 -287292683618793826013052599541333984 \\
 -13768159263526427437128637337477040 \\
 228565888652134052234795741545296 \\
 77614095705972461101015836512688 \\
 3520960278252864643823054097714 \\
 -97356045409654790379642400775 \\
 -3896041249227795478575624541 \\
 3771350531721559074253693217 \\
 1046432219183518743827391855 \\
 230951034077300006360814220 \\
 46540598189766295411264050
\end{array}
\right)
\end{equation}
\begin{displaymath}
\left(
\begin{array}{c}
 8312454155170970644087050 \\
 1303216963785678085486320 \\
 179484069218701889270145 \\
 21704356449181976554215 \\
 2299019613843658502445 \\
 212567809949303078655 \\
 17076597186282818970 \\
 1184622680244012900 \\
 70388879133182280 \\
 3544333411554144 \\
 149113361559360 \\
 5141635238784 \\
 141440701952 \\
 2982935296 \\
 45253120 \\
 439296 \\
 2048
\end{array}
\right)
\end{displaymath}
and $\kappa=12$,
\begin{equation}
\left(
\begin{array}{c}
 29240315817392452748492564980040908800000000 \\
 33830315829357716648177866950433062912000000 \\
 18120129848373902187902521023513958348800000 \\
 5915289116881436164243094987941092480000000 \\
 1300158002309897005669893560164992806400000 \\
 199467829529473754135273564780187569164800 \\
 21138278805559013390911138091825622035520 \\
 1413068924132300889731933576834973364800 \\
 35638858522310193185185202018465361456 \\
 -3080272107483898872380758495919446848 \\
 -325601356962501335852286165697351428 \\
 -5934234591428850232759442192641548 \\
 863269405306298197307322635081845 \\
 73444447458140630272343776668948 \\
 7520370028138557798308974457658 \\
 2019677853238030375854720795648 \\
 482836144179289283121850076439 \\
 96138633976278230621072341080
\end{array}
\right)
\end{equation}
\begin{displaymath}
\left(
\begin{array}{c}
 16899602534589161492520475280 \\
 2646502086250592153436542580 \\
 368329904338448688861367395 \\
 45494778945843226030218660 \\
 4981174263184321953787410 \\
 482464252353850997987160 \\
 41215604000870347678545 \\
 3093092846497823802480 \\
 202871325800172218760 \\
 11552186627193516480 \\
 566265590449861104 \\
 23631870823807488 \\
 827575382720768 \\
 23849804402688 \\
 550461669120 \\
 9776443392 \\
 125356032 \\
 1032192 \\
 4096
\end{array}
\right)
\end{displaymath}
(We are presently attempting extensions to the cases $\kappa=13, 14$.)
The leading (highest-order) coefficients in the twelve sets of two-rebit results immediately above are expressible in descending order as 
\begin{equation}
C_{3 \kappa+1}=2^\kappa; \hspace{.1in} C_{3 \kappa}=3 \times 2^{\kappa-1} \kappa (\kappa+2);\hspace{.1in} C_{3 \kappa-1}=2^{\kappa -3} \kappa  (\kappa  (\kappa  (9 \kappa +32)+24)-45);
\end{equation}
\begin{equation}
 C_{3 \kappa-2}=
2^{\kappa -4} \kappa  \left(\kappa  \left(\kappa  \left(\kappa  \left(9 \kappa ^2+42
   \kappa +52\right)-119\right)-52\right)-60\right).
\end{equation}

From these four formulas, we are able to reconstruct ($\kappa=1$) all four entries in the first column of (\ref{CoefficientArray1}). Thus, it appears that, in general, $C_{3 \kappa-i}$ is a polynomial in $\kappa$ of degree
$2 (i+1)$. (For $i=3 \kappa -1$, we obtain the constant term, of strong interest. With the knowledge of only this term, and none of the other 
coefficients, we can obtain 
$\langle |\rho^{PT}|^\kappa \rangle_{2-rebit/HS}$.)
Further, we have found that 
\begin{equation}
 C_{3 \kappa-3}= 
\end{equation}
\begin{displaymath}
\frac{1}{5} 2^{\kappa -7} (\kappa -1) \left(135 \kappa ^7+855 \kappa
   ^6+1895 \kappa ^5-1771 \kappa ^4-3091 \kappa ^3-7731 \kappa ^2+32394
   \kappa \right),
\end{displaymath}
and 
\begin{equation}
 C_{3 \kappa-4}= \frac{1}{5} 2^{\kappa -8} (\kappa -1) \kappa
\end{equation}
\begin{displaymath}
\kappa  (\kappa  (\kappa  (\kappa  (\kappa  (3 \kappa  (3 \kappa  (9 \kappa
   +59)+377)-2887)-2295)-10535)+112240)-181492)+436720.
\end{displaymath}

The numerators of our four sets ($\kappa =1, 2, 3, 4$) of two-{\it qubit} results are expressible, in similar fashion, as 
\begin{equation}
\left(
\begin{array}{cccc}
 -42 & 10944 & -6929280 & 9247219200 \\
 -1 & 4260 & -3684384 & 6039653760 \\
 6 & 220 & -456948 & 1342859616 \\
 1 & 45 & 80168 & 64072440 \\
 - & 67 & 27783 & -13235252 \\
 - & 15 & 5373 & 1080858 \\
 - & 1 & 1458 & 1160375 \\
 - & - & 282 & 278478 \\
 - & - & 27 & 50991 \\
 - & - & 1 & 7542 \\
 - & - & - & 749 \\
 - & - & - & 42 \\
 - & - & - & 1
\end{array}
\right)
\end{equation}
Of course, the leading coefficients $C_{3 \kappa +1}$ of all four numerators are 1, so they are {\it monic} in character,  
while the next-to-leading coefficients  fit the pattern 
$C_{3 \kappa}  = 3 \kappa (\kappa+3)/2$.

With our expanded computations--pursuing the Cholesky-decomposition ansatz of Dunkl--we are now able to quite substantially extend the 
sequence (\ref{sequence}) to 
\begin{equation} \label{sequence2}
\frac{\langle (|\rho| |\rho^{PT}|)^k \rangle_{2-rebit/HS}}{\langle |\rho|^{2 k} \rangle_{2-rebit/HS}} =
\end{equation}
\begin{displaymath}
\left\{0,\frac{77}{54},\frac{24}{55},\frac{209}{175},\frac{598}{833},\frac{3929}{3724},
   \frac{8432}{9867},\frac{9513091}{9555975},\frac{193880}{211497},\frac{23471937}{2408
   8922},\frac{1880}{1989},\frac{2205654099}{2276223313}\right\} \approx
\end{displaymath}
\begin{displaymath}
\{0, 1.4259, 0.43636, 1.1943, 0.71789, 1.0550, 0.85457, 0.99551, \
0.91670, 0.97439, 0.94520, 0.96900\}.
\end{displaymath}
Further, we now have for the two-qubit analogue of this sequence,
\begin{equation} \label{sequence3}
\frac{\langle (|\rho| |\rho^{PT}|)^k \rangle_{2-rebit/HS}}{\langle |\rho|^{2 k} \rangle_{2-qubit/HS}} =\left\{-\frac{3}{2},\frac{31}{10},-\frac{839}{490},\frac{3559}{1260} \right\},
\end{equation}
\begin{displaymath}
\approx \{-1.5, 3.1, -1.71224, 2.8246\}
\end{displaymath}

It is also evident at this point, in striking analogy to the general two-rebit formula (\ref{generalRebit}), that in the two-qubit scenario,
\begin{equation} \label{denominator2}
\langle |\rho|^k |\rho^{PT}|^\kappa \rangle_{2-qubit/HS}= 
\frac{\hat{A}_\kappa}{\hat{B}_\kappa} \langle |\rho|^k \rangle_{2-qubit/HS},
\end{equation}
where, again, both the numerator $\hat{A}_\kappa$ and the denominator $\hat{B}_\kappa$ are $3 \kappa$-degree polynomials in $k$, and 
(cf. (\ref{denominator}))
\begin{equation}
\hat{B}_\kappa= 2^{6 \kappa } \left(k+\frac{9}{2}\right)_{\kappa } \left(2
   k+\frac{17}{2}\right)_{2 \kappa }.
\end{equation}

In the course of this work, Charles Dunkl further communicated to us a 
result (following his joint work with K. {\.Z}yzckowski reported in \cite{Dunkl}, where 
"the machinery for producing densities from moments of Pochhammer type" was developed)
giving the univariate probability distribution over $t \in [0,1]$ that reproduces the Hilbert-Schmidt moments of $t=2^8 |\rho|$, where $\rho$ is a generic two-rebit density matrix. (It would be interesting to try to extend the methodology employed to the two-qubit and other higher-order cases. Dunkl commented that "The formula is slightly misleading near $t=1$, there the density is $(1-t)^{\frac{7}{2}}$ times an analytic function, I imagine a polynomial approximation is better for computation there, but it's obviously the stuff near zero that's important.") This probability distribution took the 
form (cf. \cite{csz}[eq. (4.3)])
\begin{equation} \label{marginaldistribution}
\frac{63}{8} \left(\sqrt{1-\sqrt{t}} \left(-8 t-9 \sqrt{t}+2\right)+15 t
   \log \left(\sqrt{1-\sqrt{t}}+1\right)-\frac{15}{4} t \log (t)\right)
\end{equation}
(see Appendix below for further details).

Of course, one may also consider issues analogous to those discussed above
for bipartite quantum systems of higher dimensionality. To begin such a course of analysis, we have found 
for the generic real $6 \times 6$ ("rebit-retrit") density matrices (occupying a 20-dimensional space) the result
\begin{equation} \label{rebitretrit}
\langle |\rho|^k |\rho^{PT}| \rangle_{rebit-retrit/HS}= 
\frac{4 k^5+40 k^4+95 k^3-220 k^2-1149 k-1170}{576 (k+4) (3 k+11) (3
   k+13) (6 k+23) (6 k+25)}
 \langle |\rho|^k \rangle_{rebit-retrit/HS}.
\end{equation}
Increasing the parameter $\kappa$ from 1 to 2, we obtained that the rational function adjustment factor for $\langle |\rho|^k |\rho^{PT}|^2 \rangle_{rebit-retrit/HS}$ is the ratio of
\begin{equation}
16 k^9+336 k^8+2616 k^7+8496 k^6+12069 k^5+101979 k^4+903539 k^3+3316809
   k^2+5620320 k+3715740
\end{equation}
to another ninth-degree polynomial 
\begin{equation}
331776 (k+5) (3 k+11) (3 k+13) (3 k+14) (3 k+16) (6 k+23) (6 k+25) (6
   k+29) (6 k+31).
\end{equation}

Additionally, for the generic complex $6 \times 6$ (qubit-qutrit) density matrices (occupying a 35-dimensional space), we have obtained the result
\begin{equation} \label{qubitqutrit}
\langle |\rho|^k |\rho^{PT}| \rangle_{qubit-qutrit/HS}= 
\frac{k^5+15 k^4+37 k^3-423 k^2-2558 k-3840}{72 (2 k+13) (3 k+19) (3
   k+20) (6 k+37) (6 k+41)}
 \langle |\rho|^k \rangle_{qubit-qutrit/HS}.
\end{equation}
\appendix
\section{Derivation by C. Dunkl of probability distribution (\ref{marginaldistribution}) over $t \in [0,1]$ having the moments of 
$t=2^8 |\rho|$}
Notes on moments, etc. C. Dunkl 4/11/11

Cholesky decomposition:

Let $C$ be a real upper-triangular $N\times N$ matrix, entries $c_{ij},$
$c_{ij}=0$ for $i>j$ and $c_{ii}\geq0$ for all $i$. Let $P=C^{t}C$, entries
$p_{ij}=\sum_{k=1}^{N}c_{ki}c_{kj}=\sum_{k=1}^{\min\left(  i,j\right)  }%
c_{ki}c_{kj}$. Consider the Jacobian matrix $\frac{\partial p}{\partial c}$
where the dependent variables are $p_{ij},i\leq j$. Claim:%
\[
\left\vert \det\frac{\partial p}{\partial c}\right\vert =2^{N}\prod_{i=1}%
^{N}c_{ii}^{N+1-i}.
\]
Lemma: suppose $y_{i}=f_{i}\left(  x_{1},x_{2},\ldots,x_{i}\right)  $, $1\leq
i\leq N$ then the matrix $\left(  \frac{\partial y_{i}}{\partial x_{j}%
}\right)  $ is lower-triangular ($0$ for $j>i$) and $\det\left(
\frac{\partial y_{i}}{\partial x_{j}}\right)  =\prod_{i=1}^{N}\frac{\partial
y_{i}}{\partial x_{i}}$.

This applies to Cholesky: order the variables: $c_{11},c_{12},\ldots
,c_{1N},c_{22},\ldots$ , $c_{2N}$, $c_{33}$, \ldots$c_{N-1,N-1},c_{N-1,N}%
,c_{NN}$. For $i\leq j$, $p_{ij}=\sum_{k=1}^{i-1}c_{ki}c_{kj}+c_{ii}c_{ij}$;
the lemma shows
\[
\left\vert \det\frac{\partial p}{\partial c}\right\vert =\prod_{i=1}^{N}%
\prod_{j=i}^{N}\frac{\partial p_{ij}}{\partial c_{ij}}=\prod_{i=1}^{N}\left(
2c_{ii}^{N-i+1}\right)  .
\]

Consider random variables, values in $0\leq t\leq1$. Moments for the Beta
distribution: let $\alpha,\beta>0$ then%
\[
\frac{1}{B\left(  \alpha,\beta\right)  }\int_{0}^{1}t^{n}t^{\alpha-1}\left(
1-t\right)  ^{\beta-1}dt=\frac{\left(  \alpha\right)  _{n}}{\left(
\alpha+\beta\right)  _{n}},n=0,1,2,\ldots.
\]
Consider $D=\det P$ where $P$ is a random positive definite $4\times4$ matrix,
trace 1. We have (expectation) for $n=0,1,2,\ldots$%
\begin{align*}
E\left(  D^{n}\right)   &  =\frac{\left(  \frac{5}{2}\right)  _{n}\left(
2\right)  _{n}\left(  \frac{3}{2}\right)  _{n}\left(  1\right)  _{n}}{\left(
10\right)  _{4n}}=\frac{2^{-2n}\left(  4\right)  _{2n}2^{-2n}\left(  2\right)
_{2n}}{2^{4n}\left(  5\right)  _{2n}\left(  \frac{11}{2}\right)  _{2n}}\\
&  =\frac{1}{2^{8n}}\frac{\left(  4\right)  _{2n}\left(  2\right)  _{2n}%
}{\left(  5\right)  _{2n}\left(  \frac{11}{2}\right)  _{2n}}.
\end{align*}
Let $X=2^{8}D$; $X$ is (equidistributed as) the product of two independent
random variables $X_{1},X_{2}$ with%
\begin{align*}
E\left(  X_{1}^{n}\right)   &  =\frac{\left(  4\right)  _{2n}}{\left(
5\right)  _{2n}}=\frac{4}{4+2n}=\frac{2}{2+n},\\
E\left(  X_{2}^{n}\right)   &  =\frac{\left(  2\right)  _{2n}}{\left(
\frac{11}{2}\right)  _{2n}}.
\end{align*}
Clearly $X_{1}$ has the density $f_{1}\left(  t\right)  =2t,0\leq t\leq1$. The
density of $X_{2}$ is%
\begin{align*}
f_{2}\left(  t\right)   &  =\frac{1}{2B\left(  2,\frac{7}{2}\right)  }\left(
1-\sqrt{t}\right)  ^{5/2},\\
\int_{0}^{1}t^{n}f_{2}\left(  t\right)  dt &  =\frac{1}{2B\left(  2,\frac
{7}{2}\right)  }\int_{0}^{1}t^{n}\left(  1-\sqrt{t}\right)  ^{5/2}dt\\
&  =\frac{1}{B\left(  2,\frac{7}{2}\right)  }\int_{0}^{1}s^{2n}s\left(
1-s\right)  ^{5/2}ds\\
&  =\frac{\left(  2\right)  _{2n}}{\left(  \frac{11}{2}\right)  _{2n}}.
\end{align*}
The density $f\left(  t\right)  $ of $X_{1}X_{2}$ is given by%
\begin{align*}
f\left(  t\right)   &  =\int_{t}^{1}f_{1}\left(  \frac{t}{s}\right)
f_{2}\left(  s\right)  \frac{ds}{s}\\
&  =\frac{2}{2B\left(  2,\frac{7}{2}\right)  }\int_{t}^{1}\frac{t}{s}\left(
1-\sqrt{s}\right)  ^{5/2}\frac{ds}{s}\\
&  =\frac{2t}{B\left(  2,\frac{7}{2}\right)  }\int_{\sqrt{t}}^{1}u^{-3}\left(
1-u\right)  ^{5/2}du\\
&  =\frac{63t}{2}\int_{\sqrt{t}}^{1}u^{-3}\left(  1-u\right)  ^{5/2}du.
\end{align*}

The integral is evaluated as follows: set $u=1-s^{2},du=-2sds$,%
\begin{align*}
f\left(  t\right)    & =63t\int_{0}^{\sqrt{1-\sqrt{t}}}\frac{s^{6}}{\left(
1-s^{2}\right)  ^{3}}ds\\
& =\frac{63t}{8}\left\{  \frac{-s\left(  15-25s^{2}+8s^{4}\right)  }{\left(
1-s^{2}\right)  ^{2}}+\frac{15}{2}\ln\frac{\left(  1+s\right)  ^{2}}{1-s^{2}%
}\right\}  _{s=0}^{s=\sqrt{1-\sqrt{t}}}\\
& =\frac{63}{8}\left\{  \left(  1-\sqrt{t}\right)  ^{1/2}\left(  2-9\sqrt
{t}-8t\right)  +15t\ln\left(  1+\sqrt{1-\sqrt{t}}\right)  -\frac{15}{4}t\ln
t\right\}  .
\end{align*}

This can be easily plotted. Also $f\left(  t\right)  =O\left(  \left(
1-t\right)  ^{7/2}\right)  $ near $t=1$.

\begin{acknowledgments}
I would like to express appreciation to the Kavli Institute for Theoretical
Physics (KITP)
for computational support in this research, and Christian Krattenthaler, Mihai Putinar, Robert Mnatsakanov, Mark Coffey and  K. {\.Z}yczkowski for various communications. Serge Provost, Jean Lasserre, Partha Biswas and Luis G. Medeiros de Souza provided guidance on reconstruction of probability distributions from moments. The earlier stages of the computations were greatly assisted by the Mathematica expertise of Michael Trott, and the later stages by the mathematical insights and suggestions of  Charles Dunkl.
\end{acknowledgments}

\bibliography{Dunkl3}

\end{document}